# A Novel Synthesis Algorithm for Reversible Circuits

Mehdi Saeedi, Mehdi Sedighi, Morteza Saheb Zamani
Quantum Design Automation Group, Computer Engineering Department
Amirkabir University of Technology
Tehran, Iran
{msaeedi, msedighi, szamani}@aut.ac.ir

*Abstract*—In this paper, a new non-search based synthesis algorithm for reversible circuits is proposed. Compared with the widely used search-based methods, our algorithm is guarantied to produce a result and can lead to a solution with much fewer steps. To evaluate the proposed method, several circuits taken from the literature are used. The experimental results corroborate the expected findings.

## 1. INTRODUCTION

An n-input, n-output, fully specified Boolean specification is called reversible if it maps each input assignment to a unique output assignment. It has been shown that using conventional irreversible logic gates leads to energy dissipation, regardless of the underlying circuit [1],[2]. Today, reversible logic design has received considerable attention in various research areas [3]-[5].

Reversible logic synthesis is defined as the ability to generate a circuit from a given reversible specification. The synthesis of reversible circuits is significantly more complex than the synthesis of traditional irreversible gates [6] and it is one of the most recent research problems.

In this paper, a non-search based synthesis method for reversible circuits is proposed. The rest of the paper is organized as follows: In Section 2, basic concepts are presented. Previous work on reversible logic synthesis is reviewed in Section 3. Our synthesis algorithm is presented in Section 4. Experimental results are reported in Section 5 and finally, Section 6 concludes the paper.

## 2. BASIC CONCEPT

An *n*-input, *n*-output gate is called reversible if it realizes a reversible function. Previously, various reversible gates with different functionalities have been proposed [7]-[9]. Among them, CNOT-based gates comprise an important class of reversible gates [10]-[18] which are also considered in this paper and denoted as follows:

Definition 1: An *n*-input, *n*-output CNOT gate $CNOT_n(x_1,x_2,...,x_n)$ passes the first *n-1* lines unchanged. These lines are referred to control lines. This gate flips the $n^{th}$ line if the control lines are all one. In other words, we have: $x_{i(out)}=x_i$ $(i<n)$, $x_{n(out)}=x_1x_2...x_{n-1} \oplus x_n$. Some authors [12] assume that complementation can also be internal to a CNOT-based gate. Therefore, it is possible to have a $CNOT_3(a',b',c)$ gate to refer to $c_{out}=c \oplus a'b'$, $a_{out}=a$ and $b_{out}=b$.

In the following section, previous algorithms for reversible circuit synthesis are reviewed.

## 3. PREVIOUS WORK

Several algorithms have recently been proposed to synthesize a reversible circuit. Toffoli in [9] presented an algorithm to implement a function using CNOT-based gates. In [10], a new incremental approach was presented using shared binary decision diagrams for representing a reversible specification and measuring circuit complexity. Some authors used transformation-based methods to optimize the synthesized results of other algorithms [11]-[13].

The authors of [14] investigated a number of techniques to synthesize optimal and near-optimal reversible circuits that require little or no temporary storage. They also provided some properties about even and odd permutation functions. As the size of a reversible circuit can be large, a practical algorithm for reversible circuit synthesis may become extremely difficult.

Due to the lack of a systematic method, search-based algorithms are widely used for reversible circuit synthesis where an extensive exploration is required to find a possible implementation of the circuit (for example see [15]-[17]). In order to guide the search process, the authors of [18] and [19] considered the use of spectral techniques to select the best possible candidate based on a predefined cost function. However, as search-based algorithms evaluate all possible gates to find an implementation of the circuit, they cannot be used to synthesize large functions.

In the following section, we propose a non-search based synthesis algorithm for reversible circuits which produces a solution for a given specification without evaluation of all possible gates during each step.

## 4. SYNTHESIS ALGORITHM

Based on the definition of a reversible specification, it can be said that a reversible Boolean specification of size *n* maps the set of integers $\{0, 1... 2^n-1\}$ onto itself probably with different order where the $j^{th}$ integer represents the $j^{th}$ minterm. For example, the reversible specification shown in Fig. 1-a may be represented as the set of integers $\{2,7,0,1,6,3,4,5\}$.

In this paper, the $i^{th}$ input (output) variable is denoted as $a_i$ ($f_i$). In addition, a general reversible specification of size *n* is shown as $F(a_1,a_2,...,a_n)=(f_1,f_2,...,f_n)$. Assume that a set of CNOT-based gates $(g_1,g_2, ...g_k)$ is used to produce $f_i$ ($i=1,...,n$) from its corresponding $a_i$ as shown in Fig. 1-b. Since the circuit is reversible, one can use the same set of gates in the reverse order, i.e. $(g_k,g_{k-1}, ...g_1)$, to produce $a_i$ from $f_i$.

| $a_1$ | $a_2$ | $a_3$ | $f_1$ | $f_2$ | $f_3$ |
|---|---|---|---|---|---|
| 0 | 0 | 0 | 0 | 1 | 0 |
| 0 | 0 | 1 | 1 | 1 | 1 |
| 0 | 1 | 0 | 0 | 0 | 0 |
| 0 | 1 | 1 | 0 | 0 | 1 |
| 1 | 0 | 0 | 1 | 1 | 0 |
| 1 | 0 | 1 | 0 | 1 | 1 |
| 1 | 1 | 0 | 1 | 0 | 0 |
| 1 | 1 | 1 | 1 | 0 | 1 |

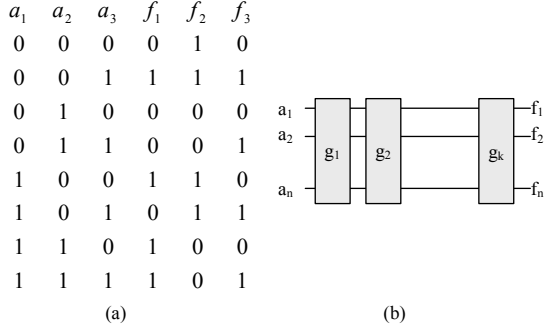

Figure 1. (a) A possible reversible specification of size 3, (b) producing $n$ reversible functions from $k$ reversible gates

Definition 2: The application of a reversible CNOT-based gate at the output side of a reversible specification $F$ is called *"output translation"*. Therefore, after using several output translations each output variable $f_i$ will be transformed to its corresponding $a_i$.

As each output translation is a reversible gate, the result of using an output translation on a reversible circuit will also be reversible. Furthermore, by using an output translation only one output variable (i.e. the last one) is changed and the others are left unchanged.

Lemma 1: (a) Applying an output translation to a given specification $F$ exchanges the location of $2^k$ minterm pairs where $k \le n-1$. (b) Conversely, exchanging the location of $2^{k-1}$ ($k=n-m+1$) minterm pairs with the following properties has the same result as applying an output translation $CNOT_m(f_{i_1}, f_{i_2}, ..., f_{i_{m-1}}, f_{i_m})$ to $F$ where $i_k \in (1...n)$ and $m \le n$:

- all of the $2^k$ minterms have the same value on $m-1$ particular bit locations.
- the two minterms of each pair differ only in one bit position.

Proof: (Case a): Assume that an output translation $CNOT_m(f_{i_1}, f_{i_2}, ..., f_{i_{m-1}}, f_{i_m})$ is applied to $F$ where $f_{i_k}$ for $k \in (1..m-1)$ can also be a complemented function. It can be easily verified that this output translation changes $f_{i_m}$ to $f_{i_1}f_{i_2}...f_{i_{m-1}} \oplus f_{i_m}$ where the value of $f_{i_1}f_{i_2}...f_{i_{m-1}}$ is 1 for only $2^k$ ($k=n-m+1$) minterms. As a result, by using $CNOT_m(f_{i_1}, f_{i_2}, ..., f_{i_{m-1}}, f_{i_m})$, the location of these $2^k$ minterms are changed. Moreover, it can be checked that $CNOT_m(f_{i_1}, f_{i_2}, ..., f_{i_{m-1}}, f_{i_m})$ exchanges the locations of all $2^{k-1}$ minterm pairs $m_i : f_{i_m}=1$ and $m_j : f_{i_m}=0$.

(Case b): Since there are $2^k$ ($k=n-m+1$) minterms which have the same value on their $m-1$ bits, there are $2^{k-1}$ minterm pairs each of which differs only in one bit position. Therefore, exchanging the location of these pairs has the same effect as applying an output translation $CNOT_m(f_{i_1}, f_{i_2}, ..., f_{i_{m-1}}, f_{i_m})$ where $i_k \in (1...m)$ and $m \le n$. □

Based on the previous definitions and lemma, the goal of our reversible synthesis algorithm is to generate a set of output translations with a specific order which when applied to the reversible specification $F$, generates $a_i$ from $f_i$. Fig. 2 shows our synthesis algorithm.

```
Input: A reversible specification F (a1, a2, ..., an) = (f1, f2, ..., fn)
Output: A set of reversible CNOT-based gates which when applied to F
produces an identity function.
Notation: The i^th function (variable) of j^th minterm is denoted as f_{i,mj}
(a_{i,mj}). Consequently, The i^th minterm of j^th function (variable) is
denoted as m_{i,fj} (m_{i,aj}).

i = 1;
repeat
  reset all of the 2^n minterms to be unvisited.
  for each minterm m_j (j = 1 ... 2^n) do
    if m_j is not visited then
      if f_{i,mj} ≠ a_{i,mj} then
      begin
        mark the minterm m_{j,fi} as a visited minterm
        select the minterm m_{k,fi} which differs from m_{j,fi} in its i^th variable
        if m_{k,fi} is below m_{j,fi} then
          exchange the locations of m_{j,fi} and m_{k,fi}. (Therefore f_{i,mj} =a_{i,mj})
          mark the minterm m_{k,fi} as a visited minterm
        else if m_{k,fi} is above than m_{j,fi} then
          if f_{p,mk} ≠ a_{p,mk} (p=1...n) for at least one p then
            exchange the locations of m_{j,fi} and m_{k,fi}. (Therefore f_{i,mj} =a_{i,mj})
            mark the minterm m_{k,fi} as a visited minterm
      end
  Extract the set of output translations (gates) based on Lemma 1
  i = (i+1) mod n;
until f_i=a_i for each i∈(1...n)
```

Figure 2. Our synthesis algorithm

The following example explains the proposed algorithm in more details:

Example 1: Consider a reversible specification $F(a_1,a_2,a_3)=(0,1,2,3,7,5,6,4)$ defined as the first and the second columns of Fig. 3.

| | | | $F$ | | | Step1 | | | Step2 | | | Step3 | | |
|---|---|---|---|---|---|---|---|---|---|---|---|---|---|---|
| $a_1$ | $a_2$ | $a_3$ | $f_1$ | $f_2$ | $f_3$ | $f_1$ | $f_2$ | $f_3$ | $f_1$ | $f_2$ | $f_3$ | $f_1$ | $f_2$ | $f_3$ |
| 0 | 0 | 0 | 0 | 0 | 0 | 0 | 0 | 0 | 0 | 0 | 0 | 0 | 0 | 0 |
| 0 | 0 | 1 | 0 | 0 | 1 | 0 | 0 | 1 | 0 | 0 | 1 | 0 | 0 | 1 |
| 0 | 1 | 0 | 0 | 1 | 0 | 0 | 1 | 0 | 0 | 1 | 0 | 0 | 1 | 0 |
| 0 | 1 | 1 | 0 | 1 | 1 | 0 | 1 | 1 | 0 | 1 | 1 | 0 | 1 | 1 |
| 1 | 0 | 0 | 1 | 1 | 1 | **1** | **0** | **1** | **1** | **0** | **0** | 1 | 0 | 0 |
| 1 | 0 | 1 | 1 | 0 | 1 | **1** | **1** | **1** | 1 | 1 | 1 | **1** | **0** | **1** |
| 1 | 1 | 0 | 1 | 1 | 0 | 1 | 1 | 0 | 1 | 1 | 0 | 1 | 1 | 0 |
| 1 | 1 | 1 | 1 | 0 | 0 | 1 | 0 | 0 | **1** | **0** | **1** | **1** | **1** | **1** |

Figure 3. The specification of Example 1 before and after three translations

Step 1: Select the first variable (i.e. $a_1$). It can be verified that the minterms of $a_1$ are placed at their right positions. Therefore, $i$ should be incremented to select the second variable (i.e. $a_2$). In addition, it can be seen that the first four minterms of $a_2$ are also positioned correctly. So, set $j=5$ and check the 5th minterm of $f_2$ (i.e. 1) and $a_2$ (i.e. 0). As these minterms are not equal, the 6th minterm of $F$ (i.e. 101) should be selected. Note that the 5th and the 6th minterms differ only in their second variable. Furthermore, the minterm *101* (the 6th minterm) is below *111* (the 5th minterm) in $F$. Therefore, these two minterms are exchanged by the algorithm. Then, set $j=7$ as the $6^{th}$ minterm has been visited previously. However, the $7^{th}$ minterm was also placed correctly which leads to set $j=8$ to verify the last minterm of $a_2$. Note that the second variable of this minterm is wrong. However, correcting it needs to change the location of the $7^{th}$ minterm (i.e. 110) which is at its right position above the $8^{th}$ minterm. Therefore, the algorithm does nothing to and goes to the next step. The third column in Fig. 3 shows the specification of $F$ after this translation.

Step 2: Select the third variable (i.e. $a_3$). Start with $j=5$ as the first four minterms were placed correctly. It can be checked that the $5^{th}$ minterm of $f_3$ (i.e. 1) and $a_3$ (i.e. 0) are not equal. Therefore, the $8^{th}$ minterm of $F$ (i.e. 100) should be selected. Note that these two minterms differ only in their third variable. Furthermore, the minterm $100$ (the $8^{th}$ minterm) is below 101 (the $5^{th}$ minterm) in $F$. Therefore, these two minterms are exchanged. Other minterms of $f_3$ (i.e. $j=6$ and $7$) are left unchanged. The forth column in Fig. 3 shows the specification of $F$ after this translation.

Step 3: Select the second variable (i.e. $a_2$) as the first variable needs no consideration (see Step 1). Based on the previous two steps and the proposed algorithm, it can be easily checked that the locations of the $6^{th}$ and the $8^{th}$ minterms are exchanged. Since after the third translation, we have $f_i=a_i$ for each $i \in (1...n)$, the algorithm is finished. The fifth column in Fig. 3 shows the specification of $F$ after this translation.

In order to find the CNOT-based implementation of each output translation, one can use the results of the previous translation and Lemma 1 to find each gate. Fig. 4 shows the implementation of each output translation and the final circuit.□

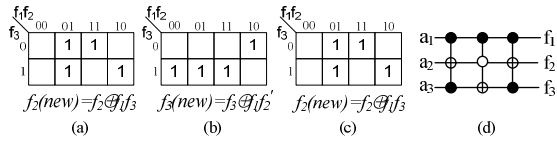

Figure 4.  The CNOT-based implementation of each output translation for (a) Step 1, (b) Step 2, (c) Step 3 and (d) the final circuit

As each output translation changes only one $f_i$ $(i=1...n)$, if the previous output translation placed the minterms of the $i^{th}$ variable at their right locations, the current output translation applied to another variable would not change their locations. Furthermore, it is important to note that the result of applying the proposed algorithm is a set of CNOT-based gates which should be applied in the reverse order to the input variables $(a_1,a_2,...,a_n)$ to produce the outputs $(f_1,f_2,...,f_n)$.

Theorem 1: The proposed algorithm will converge to a possible implementation after several steps.

Proof: Consider a reversible specification $F$ of size $n$. Assume that after the $i^{th}$ step, several minterms which are represented as a set $\Sigma$, are placed at their right positions and in the $(i+1)^{th}$ step, the algorithm works on the $k^{th}$ variable $(k \leq n)$. Suppose that the $k^{th}$ variable of a minterm, i.e. the $m^{th}$ ($m \notin \Sigma$) minterm, is not correct. Accordingly, the algorithm finds a minterm placed at location $p$ which differs from the $m^{th}$ minterm only in its $k^{th}$ variable. If $p \in \Sigma$ and $p<m$, the algorithm does nothing to avoid instability in minterm locations. However, as the $m^{th}$ minterm is placed at a wrong position (for example, the position of the $q^{th}$ minterm, $q \notin \Sigma$), there must be another minterm, i.e. the $q^{th}$ minterm, which should be exchanged with the $m^{th}$ minterm during the next steps. Therefore, the algorithm does not finish at the current step and the algorithm will reach the other cases, i.e. $p \notin \Sigma$ or $p \in \Sigma$ and $p>m$). For these cases, the algorithm exchanges the location of the $p^{th}$ minterm with that of the $m^{th}$ minterm. Then, the $k^{th}$ variable of the $m^{th}$ minterm will be correct and the algorithm moves forward to check other minterms. As each output translation does not change the results of the previous ones, the algorithm will gradually place all minterms at their right positions. Therefore, the algorithm will lead to a valid result.□

By the previous theorem, we show that the proposed algorithm will converge after several steps. In order to compare the time complexity of our proposed approach with the search based methods assume that a possible CNOT-based implementation of a reversible specification $F$ of size $n$ needs at most $h$ gates. It can be verified that $n \times 2^{n-1}$ possible gates must be evaluated to simplify $F$ at each step of a search-based method. Therefore, a search-based algorithm evaluates $(n \times 2^{n-1})^h$ or $O(n \times 2^n)^h$ gates in the worst case. On the other hand, the proposed method considers $h$ output translations, i.e. gates, in the worst case where for each translation at most $2^n$ minterms are considered. As a result, the time complexity of our algorithm is $O(h \times 2^n)$.

Compared with the search-based methods [15]-[17], the proposed algorithm needs much fewer steps to synthesize a given specification. In the following section, the experimental results are shown.

## 5. EXPERIMENTAL RESULTS

Our proposed algorithm was implemented in C++. Due to lack of space, we only maintained CNOT gate control and target lines. For example, we use $(a,b,c)$ meaning $CNOT_3(a,b,c)$. To evaluate the proposed method, we use the examples of [17]. The results of using our method and two previous search-based algorithms [15], [17] are shown in Table I. As shown in this table, the proposed algorithm not only has the ability to produce a result for all of the attempted specifications but also can reach the result with much fewer steps.

## 6. CONCLUSIONS

In this paper, a new non-search based synthesis algorithm was proposed which requires fewer steps to synthesize a given reversible specification. In order to evaluate the algorithm, we used several examples taken from the literature. It was shown that the proposed approach can lead to a result for all of the circuits very fast. The natural next step for future work seems to be working on the improvement of the resulting synthesized circuits possibly by combining the proposed approach and the search-based methods. Efforts to reach this goal are under way.

TABLE I. THE RESULTS OF USING THE PROPOSED SYNTHESIS METHOD VERSUS TWO SEARCH-BASED ALGORITHMS

| Circuit # | | Specification | Number of Gates | | Number of Searched Nodes [15][17] & Steps (our algorithm) | | | Our Synthesized Circuits |
|---|---|---|---|---|---|---|---|---|
| | | | *Our Algorithm* | *[15], [17]* | *Our Algorithm* | *[17]* | *[15]* | |
| [17] | 1 | (1,0,3,2,5,7,4,6) | 6 | 4 | 48 | 15 | 11 | $(f_1,f_2),(f_3),(f_1,f_2),(f_1,f_2,f_3),(f_1,f_3,f_2),(f_1,f_2,f_3)$ |
| | 2 | (7,0,1,2,3,4,5,6) | 3 | 3 | 24 | 300 | 761 | $(f_2,f_3,f_1),(f_3,f_2),(f_3)$ |
| | 3 | (0,1,2,3,4,6,5,7) | 3 | 3 | 24 | 10 | 7 | $(f_1,f_3,f_2),(f_1,f_2,f_3),(f_1,f_3,f_2)$ |
| | 4 | (0,1,2,4,3,5,6,7) | 7 | 5 | 56 | 786 | 156 | $(f_2,f_3,f_1),(f_1,f_2),(f_1,f_3),(f_2,f_3,f_1),(f_1,f_3,f_2),(f_1,f_2,f_3),(f_1,f_3,f_2)$ |
| | 5 | (0,1,2,3,4,5,6,8,7,9,10,11,12,13,14,15) | 15 | 7 | 240 | 8256 | 9515 | $(f_2,f_3,f_4,f_1),(f_1,f_3,f_4,f_2),(f_1,f_3,f_4,f_2),(f_1,f_2,f_4,f_3),(f_1,f_2,f_4,f_3),$ $(f_1,f_2,f_3,f_4),(f_1,f_2,f_3,f_4),(f_2,f_3,f_4,f_1),(f_1,f_3,f_4,f_2),(f_1,f_2,f_4,f_3),$ $(f_1,f_2,f_4,f_3),(f_1,f_2,f_3,f_4),(f_1,f_3,f_4,f_2),(f_1,f_2,f_4,f_3),(f_1,f_3,f_4,f_2)$ |
| | 6 | (1,2,3,4,5,6,7,0) | 3 | 3 | 24 | 4 | 4 | $(f'_2,f_3,f_1),(f_3,f_2),(f_3)$ |
| | 7 | (1,2,3,4,5,6,7,8,9,10,11,12,13,14,15,0) | 4 | 4 | 64 | 5 | 5 | $(f'_2,f'_3,f'_4,f_1),(f'_3,f'_4,f_2),(f'_4,f_3),(f_4)$ |
| | 8 | (0,7,6,9,4,11,10,13,8,15,14,1,12,3,2,5) | 3 | 4 | 48 | 139 | 230 | $(f_4,f_1),(f_3,f_2),(f_4,f_3)$ |
| | 9 | (3,6,2,5,7,1,0,4) | 8 | 7 | 64 | 66 | - | $(f_2,f_3,f_1),(f'_2,f_1),(f_2),(f_2,f_3),(f_1,f_3,f_2),(f_1,f_2,f_3),(f_2,f_3,f_1),(f_1,f_2,f_3)$ |
| | 10 | (1,2,7,5,6,3,0,4) | 8 | 6 | 64 | 77 | - | $(f_3,f_1),(f_2,f_3,f_1),(f_2),(f_1,f_2,f_3),(f_1,f_2,f_3),(f_2,f_3,f_1),(f'_1,f_3,f_2),$ |
| | 11 | (4,3,0,2,7,5,6,1) | 8 | 7 | 64 | 4387 | - | $(f_2,f_3,f_1),(f_1,f_3,f_2),(f_1,f_2),(f_1',f_2,f_3),(f_1,f_2,f_3),(f_2,f_3',f_1),(f_1,f_3,f_2),$ $(f_1,f_2,f_3)$ |
| | 12 | (7,5,2,4,6,1,0,3) | 6 | 7 | 48 | 352 | - | $(f_2,f_3,f_1),(f_3,f_1),(f'_1,f_3,f_2),(f_1,f_3,f_2),(f'_1,f'_2,f_3),(f_1,f_3,f_2)$ |
| | 13 | (6,2,14,13,3,11,10,7,0,5,8,1,15,12,4,9) | 23 | 15 | 368 | 678 | - | $(f'_4,f_1),(f'_2,f'_4,f_1),(f'_2,f_3,f_4,f_1),(f_3,f_4,f_2),(f_4,f_2),(f_1,f'_4,f_3),(f_1,f_4,f_3),$ $(f_1,f_2,f_4,f_3),(f_1,f_2,f_3,f_4),(f_1,f_2,f_3,f_4),(f_3,f_4,f_1),(f_2,f_4,f_1),(f_1,f_3,f_4,f_2),$ $(f_1,f_3,f_4,f_2),(f'_1,f_2,f_4,f_3),(f_1,f_2,f_3),(f_1,f_2,f_4,f_3),(f_1,f_2,f_3,f_4),$ $(f_1,f_2,f_3,f_4),(f_2,f_3,f_4,f_1),(f_1,f_3,f_4,f_2),(f_1,f_4,f_3),(f_1,f_2,f_3,f_4)$ |
| Avg. | | | 7.46 | 5.76 | 87.38 | 1159 | | |